\documentclass[aps,prl,reprint,twocolumn,showpacs,superscriptaddress]{revtex4}

\usepackage{graphicx}
\begin{document}
\title{Photoemission Spectroscopy of Magnetic and Non-magnetic Impurities on
    the Surface of the Bi$_2$Se$_3$ Topological Insulator} 

\author{T. Valla}
\email{valla@bnl.gov}
\author{Z.-H. Pan}
\affiliation{Condensed Matter Physics and Materials Science Department, Brookhaven National Lab, Upton, NY 11973}
\author{D. Gardner}
\author{Y.S. Lee}
\affiliation{Department of Physics, Massachusetts Institute of Technology, Cambridge, MA 02139}
\author{S. Chu}
\affiliation{Center for Materials Science and Engineering, Massachusetts Institute of Technology, Cambridge, MA 02139}

\date{\today}

\begin{abstract}
Dirac-like surface states on surfaces of topological insulators have a 
chiral spin structure that suppresses back-scattering and protects the 
coherence of these states in the presence of non-magnetic scatterers. 
In contrast, magnetic scatterers should open the back-scattering channel via the spin-flip processes and degrade the state's coherence.
We present angle-resolved photoemission spectroscopy studies of the 
electronic structure and the scattering rates upon adsorption of 
various magnetic and non-magnetic impurities on the surface of 
Bi$_2$Se$_3$, a model topological insulator. We reveal a remarkable 
insensitivity of the topological surface state to both non-magnetic and 
magnetic impurities in the low impurity concentration regime. 
Scattering channels open up with the emergence of hexagonal warping 
in the high-doping regime, irrespective of the impurity's magnetic moment.
\end{abstract}
\vspace{1.0cm}

\maketitle 
Topological insulators (TIs) belong to a new class of insulators in which the 
bulk gap is inverted due to the strong spin-orbit coupling. On the 
boundaries or interfaces of these materials with ordinary 
("trivial") insulators, gapless states inevitably occur, topologically 
protected by the time reversal symmetry 
\cite{Kane2005,Bernevig2006a,Konig2007}. 
Three-dimensional topological insulators have surface states with an odd 
number of massless Dirac cones in which the spin of an electron is locked 
perpendicular to its momentum in a chiral spin-structure where electrons 
with opposite momenta have opposite spins 
\cite{Fu2007a,Noh2008,Hsieh2008,Zhang2009,Hsieh2009,Chen2009,Pan2011}. 
A direct consequence of this spin-momentum locking is that a 
backscattering, which would require a spin-flip process, is not allowed if a time-
reversal-invariant perturbation, such as non-magnetic disorder, is present 
\cite{Fu2007a}. This makes topological insulators potentially very 
promising materials that could serve as a platform for spintronics and for quantum 
computing applications, where spin-coherence is crucial. In contrast, a time-reversal symmetry 
breaking perturbation, such as introduction of magnetic impurities on the surface, 
is expected to open a back-scattering channel and induce a gap at the Dirac point of the topological surface state (TSS) \cite{Fu2009,Liu2009,Zhou2009,Guo2010,Biswas2010,Chen2010a,Wray2010}. 

Even though it might be expected that these fundamental predictions would be checked very quickly, the experiments that would directly probe the sensitivity of the TSS and differentiate between the two types of disorder are still lacking. Scanning tunneling microscopy (STM) experiments have shown that backscattering is indeed strongly suppressed or completely absent, despite strong atomic scale disorder \cite{Roushan2009,ZhangSTM2009,Hanaguri2010}. In angle resolved photoemission spectroscopy (ARPES), there has been very little quantitative work on the scattering rates. One study \cite{Park2010} has indicated that the major decay channel for the TSS is scattering into the bulk states, either elastically, on defects, or inelastically, via the electron-electron interaction. More recent studies have also shown that the adsorption of various non-magnetic atomic/molecular species on the surface of a topological insulator induces electronic doping and partial filling of additional spin-orbit split states \cite{Benia2011,Bianchi2011}. It has been also suggested that magnetic impurities, both in the bulk and on the surface, open a small gap at the Dirac point \cite{Chen2010a,Wray2010}. However, the most fundamental question -  how the magnetic moment of an impurity affects the scattering - has remained unanswered. Even for non-magnetic perturbations, it would be highly desirable to know how the TSS behaves as the concentration of impurities increases and how it is affected by the presence of other states that become partially occupied by electron doping. Would the presence of such states, that come in spin-orbit split pairs, therefore allowing inter-band scattering, both with and without spin-flip, degrade the TSS\rq{} coherence? Or, would the intra-band scattering dominate? We note that none of these questions have been addressed in experiments and that there has been no systematic studies of the scattering rates on any kind of impurities. 

\begin{figure*}[htbp]
\begin{center}
\includegraphics[width=14cm]{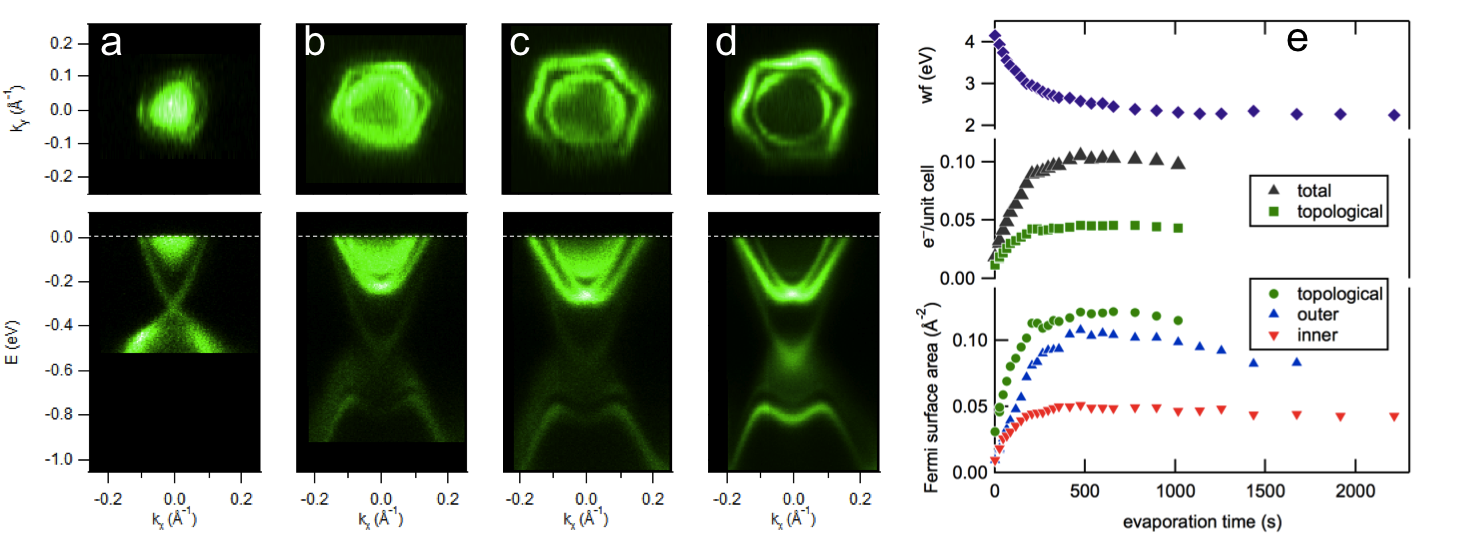}
\caption{Surface doping of Bi$_2$Se$_3$. a) to d) ARPES spectra from 
Bi$_2$Se$_3$ at various stages of Rb deposition,
 showing the Fermi surface (upper panels) and the (E,$k$) dispersion of 
photoemission intensity along the momentum line
 slightly off the $\Gamma$M line in the surface Brillouin zone (lower 
panels). a) pristine surface, b) intermediate doping 
and c) maximal doping, taken at $h\nu=21.3$ eV and at  
$h\nu=18.7$ eV (d). 
e) Fermi surface area of the TSS and of the lower Rashba doublet (bottom), 
charge doped into these states (middle) and work function (top) as functions of Rb deposition time.}
\label{Fig1}
\end{center}
\end{figure*}
Here, we present quantitative experimental studies of scattering rates on on the surface of Bi$_2$Se$_3$, a model TI. We directly compare the effects of non-
magnetic and magnetic impurities on the TSS and, quite unexpectedly, we find that there is essentially no difference between these two types of scatterers. Both the scattering and the impurity induced development of the surface electronic structure seem remarkably insensitive to the type of disorder. Instead, we find that the scattering rates are sensitive to the Fermi surface shape, which can be tuned by the doping, irrespective of the impurity\rq{}s magnetic moment. We also find no evidence for an opening 
of a gap at the Dirac point of the TSS.

The single crystal samples were synthesized by mixing stoichiometric 
amounts of bismuth and selenium with trace amounts of arsenic in evacuated quartz tubes \cite{Steinberg2010}. The ARPES experiments were carried out at the U13UB beamline of the National Synchrotron Light Source with the photons in the range between 15.5 and 22 eV. The electron analyzer was a Scienta SES-2002 with the combined energy resolution around 8 meV and the angular resolution of $\sim 0.15^\circ$. 
Samples were cleaved \textit{in-situ} in the UHV chamber with the base pressure of $3\times 10^{-9}$ Pa. Ni was deposited using an e-beam evaporator, Cu and Gd were evaporated 
from a resistively heated tungsten basket, while alkalies were deposited from 
commercial (SAES) getter sources with the samples kept at $\sim15$ K 
during the deposition and ARPES measurements. 

Figure 1 shows the development of surface electronic structure upon 
deposition of rubidium on the Bi$_2$Se$_3$ surface. The rapidly 
dispersing conical band in the pristine sample represents the 
TSS with the Dirac point around 0.32 eV below the 
Fermi level. At binding energy higher than 0.4 eV the TSS 
overlaps with the bulk valence band (BVB) and near the Fermi level, the 
bulk conducting band (BCB) is visible inside the surface state cone, indicating
 the electron doping of Bi$_2$Se$_3$ by Se vacancies. The TSS has an almost 
perfectly circular Fermi surface. Upon Rb deposition, TSS is further 
doped with electrons,  evident from the down-shift of the Dirac point and 
the growing Fermi surface that acquires a pronounced hexagonal warping. 
However, this is not the only effect of doping: new states are also being 
formed and progressively filled with electrons donated by adsorbed Rb. In 
panels c) and d) we show the stage of Rb deposition at which the 
maximal charge transfer into the surface electronic structure of 
Bi$_2$Se$_3$ is reached. At this stage, in addition to the original 
TSS, two pairs of new states are visible at lower 
binding energies. Each pair consists of two spin-orbit split states, 
displaced in momentum in a Rashba-type manner, intersecting at new Dirac 
points at the zone center. These states also have surface character as 
they do not disperse with $k_z$. At the highest doping levels, the 
outermost state  becomes almost degenerate with the TSS, forming 
the Fermi surface nearly equal in shape and size. Its 
inner counterpart is significantly smaller, retaining the perfectly 
circular Fermi surface, even at the highest doping. We also observe new valence states 
below the Dirac point of the TSS. Although their dispersion 
near the zone center resembles the dispersion of the BVB, 
the lack of $k_z$ dispersion indicates their surface 
character. 
\begin{figure}
\begin{center}
\includegraphics[width=7cm]{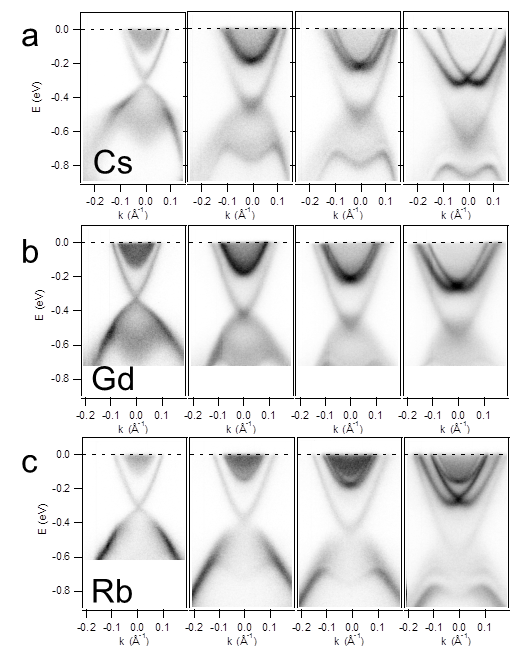}
\caption{Development of the surface electronic structure with Cesium  (a), Gadolinium (b) and Rubidium (c) adsorption at the surface of Bi$_2$Se$_3$. The spectra were
 recorded along the $\Gamma$M line in the surface Brillouin zone and at  
$h\nu=18.7$ eV for Cs and Gd and along $\Gamma$K line and at $h\nu=21.3$ eV
 for Rb. The leftmost panels correspond to the pristine surface, while 
the rightmost panels show the electronic
 structure near the maximal doping achieved with each dopand.}
\label{Fig2}
\end{center}
\end{figure}

Fig. 1e) summarizes the changes 
in some of the measured quantities with Rb doping. 
The surface doping level was determined by measuring the Fermi surface area of the TSS and of the lower Rashba-split doublet: $A_T$ (TSS), $A_O$ (outer Rashba state) and $A_I$ (inner Rashba state). The upper Rashba doublet was not taken into account. The total charge (per surface unit cell) is then $q=(A_T+A_O+A_I)/A_{BZ}$, where $A_{BZ}=2.662$ \AA$^{-2}$  represents the Brillouin zone area.
At maximal doping, nearly 0.105 $e^-$ per surface unit cell is transferred from Rb 
into the three states shown here. If the second pair of states (better 
resolved in Fig. 2c) is counted, then the total charge transfer is 
$\sim$0.14 $e^-$. The surface charge density $n=q/A_{UC}$, where $A_{UC}= 44.487$ \AA$^2$ is the area of the unit cell in real space, could be tuned from $\sim 1\times10^{12}$ cm$^{-2}$ (clean sample) to $\sim 5\times10^{13}$ cm$^{-2}$ (maximal doping). As a Rb atom can donate at most one electron, the 
measured charge transfer implies that the average Rb-Rb distance could be 
shorter than 3 unit cells. Scattering on Rb would then lead to the very short 
mean free path for surface electrons 
($\sim3$ surface unit cells). However, Fig. 1 suggests that all the 
states are still very coherent, with the mean free paths  $\ell=1/\Delta k$ 
in the range of 100 \AA, where $\Delta k$ is the 
momentum spread of the Fermi surface, measured from the momentum 
distribution curves (MDCs) \cite{Valla1999}. 
Insensitivity to impurity scattering might be expected for the 
TSS, but only in the absence of other states that could  
open the inter-band scattering channels. Thus, the retained coherence of all the detected 
states is somewhat surprising.

In Fig. 2, we compare the effects of different adsorbates on the surface 
electronic structure of Bi$_2$Se$_3$ - in particular we compare the non-
magnetic impurities, Rb and Cs, with Gd whose atoms have large magnetic 
moments, $\sim 8$ $\mu_B$. We have also studied adsorbed Ni and Cu (not 
shown). Surprisingly, there is no visible difference in 
the spectra for different adsorbates, if taken at the same photon energy. 
In the recent study where iron was deposited on Bi$_2$Se$_3$, the 
electronic structure also looks very similar \cite{Wray2010}. Relative 
intensities of the states that form the Fermi surface depend on photon 
energy, reflecting the variation of ARPES matrix elements. Thus, 
the higher pair of Rashba-split states, hardly visible in Cs and 
Gd covered surface is clearly resolved in Rb doped system, measured at 
different photon energy. However, none of these states disperse with 
$k_z$, reflecting their surface character. We also note that the maximal 
doping level achievable with different adsorbates increases from Ni to Gd 
to Cu to Rb to Cs. 

The most important observation from the spectra in 
Fig. 2 is that, contrary to the expectations, the magnetic state of the 
adsorbate does not seem to play a significant role in the scattering. At 
similar stages of doping with different adsorbates, the TSS 
seems similarly coherent. The same is true for the Rashba-
split states. Further, it appears that all the adsorbates have a similar 
effect on the spectral region around the Dirac point of TSS, with no clear gap formation. 

\begin{figure}
\begin{center}
\includegraphics[width=7.5cm]{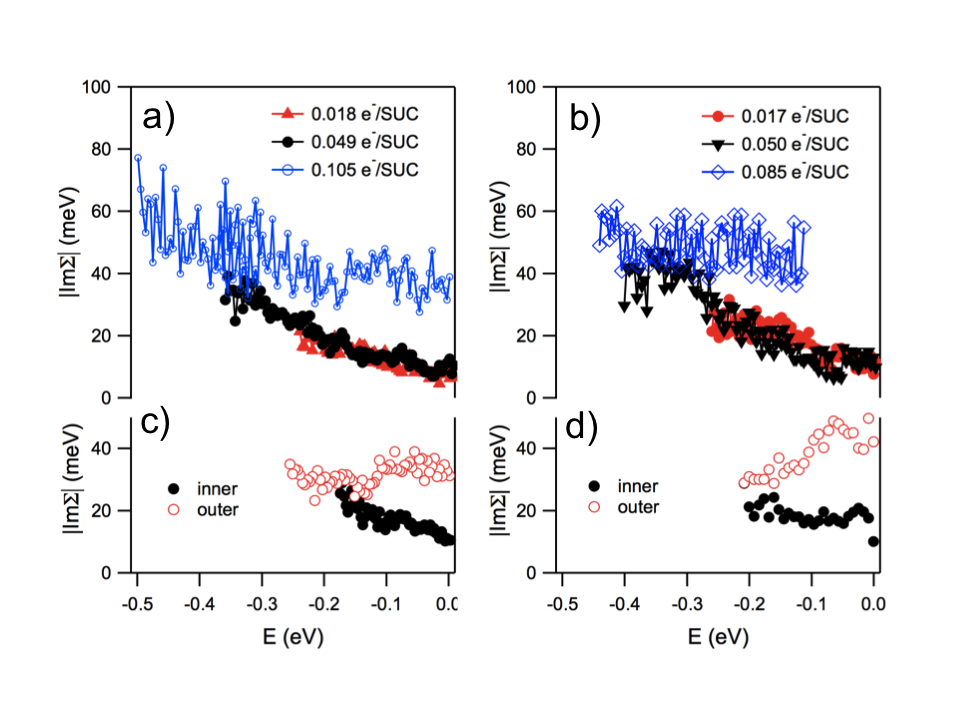}
\caption{Im$\Sigma(\omega)$ of the topological surface state upon adsorption of 
Rubidium (a) and Gadolinium (b) for several different doping levels, as 
indicated. c) and d) show Im$\Sigma(\omega)$ of the lower Rashba-split 
doublet for Rb and Gd doped surfaces, respectively, near the maximum 
doping.}
\label{Fig3}
\end{center}
\end{figure}
\begin{figure*}[htbp]
\begin{center}
\includegraphics[width=14cm]{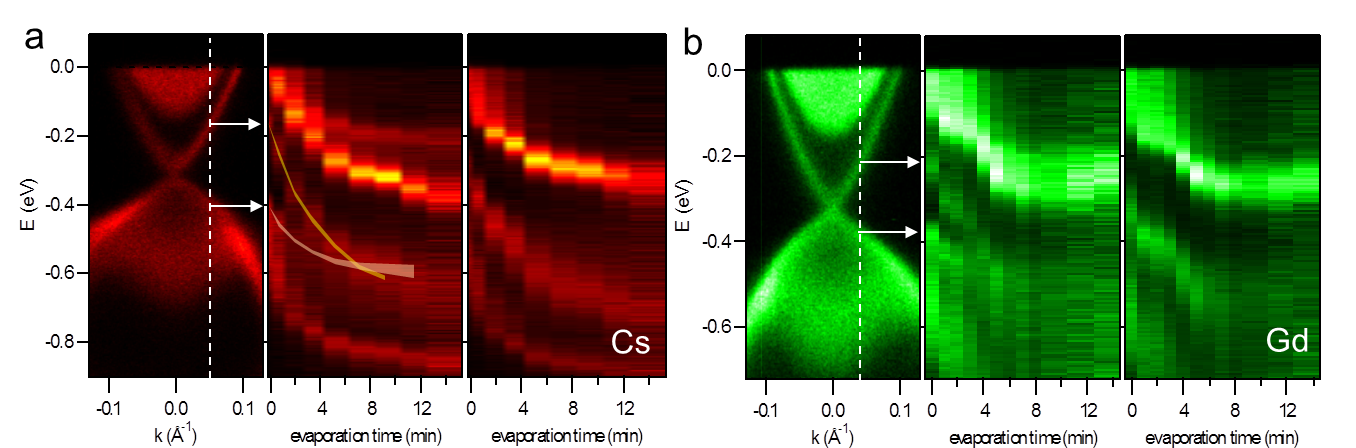}
\caption{a) Development 
of the surface electronic structure with Cs doping. Middle panel shows 
shift of the states at $k_x=0.05$ \AA, while the right panel shows the $\Gamma$ 
point in the surface BZ during Cs deposition. b) Surface electronic structure upon Gd doping. Middle (right) panel represents $k_x=0.03$ \AA ($k_x=0$) point.}
\label{Fig4}
\end{center}
\end{figure*}
In Fig. 3 we show the imaginary part of the quasi-particle self-energy, 
$|$Im$\Sigma(\omega)|=\Gamma(\omega)/2$, where $\Gamma(\omega)$ represents 
the scattering rate, as a function of binding energy for TSS for several 
different  concentrations of Rb and Gd atoms on 
the surface of Bi$_2$Se$_3$. Scattering rates are determined from 
$\Gamma(\omega)=2|$Im$\Sigma(\omega)|=\Delta k(\omega)v_0(\omega)$, where 
$\Delta k(\omega)$ is the measured full width at half maximum of the 
Lorentzian-fitted peak in MDC, and $v_0(\omega)$ is the group velocity of 
the state at energy $\omega$.  
The TSS remains very coherent until the 
concentration of adsorbed atoms reaches the level at which the Fermi 
surface becomes heavily hexagonally warped, regardless of whether 
the adsorbates are magnetic or non-magnetic. For similar doping levels, 
the scattering rates are essentially the same for Rb and Gd covered surfaces. 
Pristine surfaces and surfaces with relatively low concentration of 
impurities, show very low Im$\Sigma$ at the Fermi level, indicating long 
coherence lengths of TSS, $\ell> 150$ \AA. Even at the doping 
levels $\sim0.05$ e$^-$ per surface unit cell, where the average distance 
between the impurities is shorter than $\sim5$ unit cells, the TSS remains unaffected.
Im$\Sigma$ slowly increases with 
energy as $\propto\omega^2$, indicating that the inelastic electron-
electron scattering has a Fermi-liquid-like form.
For high impurity concentrations, $\sim0.1$ e$^-$ per surface unit cell, 
Im$\Sigma$ reaches the value of $\sim 40$  meV at the Fermi level, 
corresponding to the mean free path of $\sim 70$ \AA,  and is nearly 
energy independent. Due to the partial overlap with the significantly 
more intense outer Rashba state, we could not reliably determine the width of 
the TSS at low energies.
We also show the Im$\Sigma$ for the two states that form the lower 
Rashba-split doublet. There is a significant difference between the states forming the doublet: the outer state 
is significantly broader than the inner one and is similar in width to 
the TSS at this concentration level. This is again true for 
both magnetic and non-magnetic impurities.

In Fig. 4 we illustrate the effects of Cs and Gd deposition on the 
spectral region near the Dirac point of TSS. We 
show the spectral intensity at the point slightly displaced from the 
$k_x=0$ (middle panels) and exactly at the $k_x=0$ point (right panels), 
as a function of Cs and Gd deposition time. Contrary to the expectations, 
both metals have similar effects: with the deposition of these metals, it 
seems as if the lower and the upper parts of the Dirac cone  penetrate 
each other. Thus, at small but finite $k_x$, the two branches merge and 
possibly intersect after $\sim 6$ min of evaporation. If the gap opens at  
the Dirac point, as might be expected for magnetic impurities, the upper 
and the lower branch of the TSS cone should remain separated. 
The separation should also occur at $k_x=0$, once the gap 
opens. Our results suggest that neither of the 
adsorbates opens a clear gap at the Dirac point of the TSS. 
We also see no evidence of a gap at the second Dirac point, where the states forming the lower 
Rashba-split doublet intersect (Fig. 2). This suggests that the Kramer's points, 
i.e. the points where the spins are degenerated in the unperturbed 
system, are more robust to magnetic perturbations than expected. One 
possible reason for this insensitivity could be a strongly localized 
magnetic moment ($f$ orbitals) in adsorbed Gd, resulting in a very small 
scattering cross-section. However, similar results for adsorbed nickel 
and iron \cite{Wray2010} with the more delocalized moments, would argue against this explanation.

Our experiments show that the quasi-particle scattering on 
the surface of a TI is not affected by magnetic 
moments of impurity atoms. This might imply that the scattering 
rates are dominated by the small momentum transfer events and not by back-scattering. Then, the existence 
of multiple Fermi surfaces, allowing both the intra-band and inter-band scattering, 
and the observation that the inner Rashba-split Fermi surface is always sharper than its outer 
counterpart and the TSS might suggest that the former one has 
the opposite spin helicity than the latter two. However, recent 
calculations \cite{Wray2010} suggest that the spin helicities of these 
three states alternate (L-R-L). If this is the case, our results would imply 
that the inter-band scattering is strongly suppressed. Indeed, we do not 
see any anomalies in the scattering rates at the thresholds for the inter-band 
channels. Therefore, we could conclude that the observed 
broadening with adsorption of impurities reflects the increase in intra-
band scattering as the size and the warping of the Fermi surface grows 
with doping \cite{Chen2009,Fu2009,Zhou2009,Lee2009,Kuroda2010}. 
We note that these effects will likely play determining role in a performance of any electronic device based on a topological insulator, because any environmental doping will inevitably affect the surface 
state mobility, $\mu_S=e\ell_{tr}/(\hbar k_F)$, in transport experiments. Even though the transport mean free path, $\ell_{tr}$, might be significantly longer than $\ell$, especially when back-scattering is suppressed, our results indicate that mobilities will be reduced by the doping, implying that the full potential of TIs could only be realized in a controlled, preferably ultra-high vacuum environment, or by an inert capping of the surface.

In conclusion, we have observed that magnetic moment of an impurity does not play a dominant role in the scattering of the TSS. However, with the increasing doping, the state becomes warped, and the scattering eventually increases - irrespective of impurity's magnetic moment. Therefore, the TSS does not remain protected indefinitely, even when doped with non-magnetic impurities. Further, we have not seen any difference in the spectral region around the Dirac point between magnetic and non-magnetic adsorbates, questioning previous claims that the observed spectral features indicate a magnetism-induced gap \cite{Wray2010}.

We acknowledge valuable discussions with G. Gu, P. D. Johnson, M. Khodas, R. Konik and E. Vescovo.
The work at BNL was supported by the US Department of Energy, Office of Basic 
Energy Sciences, under contract DE-AC02-98CH10886. The work at MIT was 
supported by the US Department of Energy under Grant No. DE-FG02-07ER46134..


\end{document}